# Physicochemical Effects in Aging Aqueous Laponite Suspensions

A. Shahin and Yogesh M Joshi*

Department of Chemical Engineering, Indian Institute of Technology Kanpur, Kanpur 208016. INDIA.

* E-Mail: joshi@iitk.ac.in

**Abstract**

We study aging behavior of aqueous suspension of Laponite as a function of concentration of Laponite, concentration of salt, time elapsed since preparation of suspension (idle time) and temperature by carrying extensive rheological and conductivity experiments. We observe that temporal evolution of elastic modulus, which describes structural build-up and aging, shifts to low times for experiments carried out for higher concentration of Laponite, higher concentration of salt, greater temperature and on higher idle time while preserving the curvature of evolution in the solid regime (elastic modulus greater than viscous modulus). Consequently appropriate shifting of evolution of elastic modulus in the solid regime leads to *aging time – idle time – salt concentration – Laponite concentration– temperature superposition*. Existence of such superposition suggests generic nature of microstructure buildup irrespective of mentioned variables in the explored range. Behavior of shift factors needed to obtain the superposition indicate that energy barrier associated with structural buildup decreases with increase in idle time and temperature; and decreases linearly with increase in concentration of Laponite and that of salt. The conductivity experiments show that ionic conductivity of the suspension increases with increasing Laponite concentration, salt concentration, temperature and very importantly the idle time. We also analyze the inter-particle interactions using DLVO theory that suggests increase in idle time, temperature and salt concentration increases height of repulsive energy barrier while decreases width of the same when particles approach each other in a parallel fashion. However when particles approach each other in a perpendicular fashion, owing to dissimilar charges on edge and face, energy barrier for attractive interaction is expected to decrease with increase in idle time, temperature and salt concentration. Analysis of rheological and conductivity experiments suggest strong influence of attractive interactions on the low energy structures in aqueous suspension of Laponite.

**I. Introduction:**

Smectite clay minerals are important class of colloidal materials with immense academic importance and wide ranging applications.[1, 2] Smectite clays have oblate shape (layered like) with aspect ratio in the range 25 to 1000.[1] Particularly clays embedded in aqueous media are used in petroleum, healthcare, cosmetic, etc. industries, as rheology modifiers.[3] Therefore, it is not surprising that rheological and phase behaviors of clays have been very active areas of research over past several decades.[1, 2, 4-17] In the recent literature synthetic clay mineral called Laponite has received significant attention due to its rich physical behavior and industrial applications. In addition, aqueous Laponite suspension shows time dependence evolution of its microstructure and physical properties that is reminiscent to physical aging in molecular, spin and colloidal glasses.[18, 19] In this work we carry out extensive rheological and conductivity studies on aqueous suspension of Laponite and analyze various distinct characteristic features of aging observed in this system.

Physical aging is a natural tendency of those materials that, owing to kinetic constraints, have fallen out of thermodynamic equilibrium.[20] In a physical aging process material explores its phase space and undergoes time dependent evolution in order to progressively attain lower energy microstructures.[21] There are many soft materials such as concentrated suspensions and emulsions, cosmetic and pharmaceutical pastes, colloidal gels, clay suspensions, foams, etc. that demonstrate physical aging and have enormous academic and industrial importance.[22] Physical aging is, in principle, a reversible phenomenon. Application of deformation field reverses the effects of aging by rejuvenating the material to higher energy levels.[23, 24] Physical aging is analyzed by monitoring affected physical properties of the material that show time dependent change. In soft materials, the light scattering techniques are commonly used to probe structure and evolution of the mobility of the constituents of materials,[25-30] which also leads to the relaxation time of the same. Bulk rheological techniques are also used to monitor evolution of viscoelastic properties such as elastic and viscous modulus which get strongly affected by physical aging.[31-33] Recently microrheological techniques have also been used to analyze soft glassy materials.[34-36] For various soft materials that are thermodynamically out of equilibrium, including aqueous Laponite suspension, ample literature is available that employs scattering as well as rheological techniques to study the aging behavior. However, since effect of aging in soft glassy materials can be reversed by application of deformation field alone, rheological techniques are crucial for studying these phenomena.[18, 31, 37-39]

Laponite (hydrous sodium lithium magnesium silicate, $Na_{+0.7}[(Si_8Mg_{5.5}Li_{0.3})O_{20}(OH)_4]_{-0.7}$) is a synthetic layered silicate clay mineral available as a white powder. Laponite particle has a disk like shape with diameter in the range 25 – 30 nm with thickness of 1 nm.[40] In dry form Laponite particles are present in stacks with sodium ions residing in the inter-particle gallery. When dispersed in aqueous medium sodium ions dissociate rendering the faces of Laponite disk a permanent negative charge. The edge of Laponite particle, which is composed of hydrous oxides of magnesium and silica, is reported to acquire a positive charge at low pH (below the pH of around 11).[41] Overall Laponite particles share edge – to – face attractive interactions while face – to – face repulsive interactions among each other. Addition of salt such as NaCl enhances concentration of cations and anions in the suspension which shield the charges on the particle thereby reducing repulsion among the particles. Addition of Laponite in water, typically beyond 2 weight %, increases its viscosity and elastic modulus by several orders of magnitude over a short duration.[31] Microstructure of Laponite suspension and its time dependency that is responsible for such enormous increase in viscosity is, however, not completely understood even after more than fifteen years of research on this topic.[42-44] Issue of contention is whether microstructure of Laponite is an attractive gel (interconnected particles through positive edge – negative face contact)[16, 30, 42, 45-47] or a repulsive glass (particles in a self-suspended state in a liquid media owing to repulsion among them).[15, 48-50]

Some reports claim existence of both the states over a certain concentration regime.[51] Interestingly, several groups have observed Laponite disks to show anisotropic orientation in a suspended state.[6-9, 13] Recently our group reported that anisotropy originates at air – suspension interface and percolates into the bulk as a function of time.[52]

Aging behavior of Laponite suspension can be rheologically monitored by applying small amplitude oscillatory shear to shear melted suspension so as to record evolution of elastic ($G'$) and viscous ($G''$) modulus as a function of time.[39, 44, 53, 54] However, it has been observed that, unlike many soft materials that show complete reversal of aging when strong deformation field is applied, the evolution of viscoelastic properties in aqueous suspension of Laponite is only partly reversible over time scale of days.[19, 37, 39, 44] Shahin and Joshi[44] observed that under large amplitude oscillatory shear (shear melting protocol in an oscillatory mode), aqueous suspension of Laponite demonstrates a plateau of complex viscosity after a sufficient time of shearing, which has a higher value for samples kept idle over longer duration since preparation of suspension (we represent this time as idle time). In addition, the temporal evolution of $G'$ and $G''$ after shear melting is stopped, shifts to lower times for experiments carried out on higher idle times. It was concluded that the structure formed by Laponite particles in a suspension cannot be completely destroyed by application of shear suggesting irreversibility in aging over duration of days.[39, 44] This observation suggests shear melted sample on a higher idle time is in a more matured (low energy) state. Furthermore, for suspensions having greater concentration of salt (NaCl) or experiments carried out at greater temperature, the evolution of $G'$ is also observed to shift to lower times suggesting rate of formation of the structure to be faster with increase in the two mentioned variables.[31, 37, 44, 55] In this paper we extend this work and analyze effect of different concentrations of Laponite, different concentrations of salt for each concentration of Laponite and temperature on the aging behavior at various idle times by carrying out extensive rheological experiments. We simultaneously perform conductivity studies as a function of the same variables and estimate inter-particle interactions using DLVO theory for Laponite suspension. We believe that both these studies give complementary information leading to new insights into various features of aging dynamics in this system.

## II. Material and Experimental Protocol:

In this work we have used smectite hectorite clay Laponite RD® procured from Southern Clay Products Inc. White powder of Laponite was dried at 120°C for 4 hours before mixing with deionized water having pH 10 and predetermined amount of NaCl. The pH was maintained by addition of NaOH. Mixing was carried out using an ultra turrex drive for a period of 45 minutes. Subsequent to mixing, suspensions were stored in sealed polypropylene bottles at the room temperature. In this work we have used five Laponite concentrations: 2, 2.4, 2.8, 3.2 and 3.5 weight %. For each concentration of Laponite, suspensions were prepared with five concentrations of salt (NaCl) in the range: 0.1 to 7 mM of externally added $Na^+$ ions (concentration of 0.1 mM is achieved merely by NaOH used to maintain pH, therefore this concentration corresponds to no salt situation). We carried out the rheological, ionic conductivity and complexometric titration experiments on these 25 samples on various days elapsed since preparation of the suspensions up to 21 (idle time) days at an interval of 3 days (For some samples we also carried out experiments up to 60 days). Complexometric titration[56] experiments using EDTA were carried out to confirm stability of these suspensions at regular interval. These experiments did not show presence of $Mg^{2+}$ ions for any sample over the above mentioned period, thereby ruling out possibility of $Mg^{2+}$ leaching from the Laponite particles.[56, 57]

Rheological experiments were performed using a stress controlled rheometer AR 1000 (Couette geometry with a bob diameter of 28 mm and a gap of 1 mm). The procedure employed in the rheological experiments is described in figure 1. For every experiment, at specific idle time (mentioned in days) a fresh sample was loaded in the couette geometry using an injection syringe and proper care was taken to avoid entrapment of air bubbles. After attaining thermal equilibrium the sample was shear melted using an oscillatory stress of 60-80 Pa (varied depending on the concentration of Laponite) and a frequency of 0.1 Hz. The time at which shear melting was stopped marks beginning of aging time ($t_w$=0). Subsequent to shear melting, aging experiments were carried out by applying small amplitude oscillatory shear stress with magnitude 10 Pa at frequency of 0.1 Hz. An evolution of the viscoelastic properties of a suspension was monitored as a function of aging time ($t_w$). It should be noted that, owing to time dependent increase in elastic modulus and viscosity, rheological linear response regime undergoes continuous broadening as a function of aging time. However estimation of linear response regime poses some practical difficulties for samples having small age. When timescale associated with stress sweep experiment is larger than the age of the sample, evolution of viscoelastic properties takes place over the duration of the experiment forbidding estimation of linear response regime. At higher ages, the aging experiments were indeed carried out in the linear response regime. Interestingly, irrespective of age of the sample strain response to the stress controlled experiments was always close to harmonic justifying usage of $G'$ and $G''$ as discussed before in greater details.[39, 44, 58] For all the above mentioned concentrations, aging experiments were carried at 10°C.

We also measured ionic conductivity (Cyberscan PC 6000) of all the samples at regular intervals since preparation of suspension up to 18 days. In order to measure conductivity, suspension samples were shear melted using injection syringe with needle having 0.5 mm diameter and 30 mm length. In this process soft solid shear melts and forms liquid. Our experience suggests that, within experimental uncertainty, extent of shearing does not show any effect on conductivity. However, as expected, for low viscosity sample, conductivity meter shows a steady state over shorter duration. For significantly aged samples measurement of accurate value of conductivity becomes difficult because of high viscosity of the suspension. Moreover, under these conditions measurement of ionic conductivity as a function of temperatures becomes more difficult owing faster increase in viscosity of the suspension at high temperatures.

In addition to the experiments at different Laponite and salt concentrations, we also studied effect of temperature in the range 10 to 40°C on the evolution of viscoelastic properties of 2.8 weight % Laponite suspension with 5 concentrations (0.1 to 7 mM) of NaCl. The temperature dependence was studied over idle time duration from 3 days up to 60 days at regular intervals. In all the rheological experiments a thin layer of low viscosity silicon oil was applied on the free surface to prevent evaporation.

## III. Results:

We begin by discussing results of a rheological study. Before starting the aging experiments all the samples were shear melted until complex viscosity of the respective suspension attained a steady state plateau. This steady state plateau of complex viscosity demonstrated higher value for experiments carried out on greater idle times indicating inability of strong shear deformation field to destroy structure formed during aging irrespective of the concentration of Laponite and that of salt. Subsequently, shear melting was stopped and suspension was allowed to age. It is usually observed that shear melting carried out at different stress levels does not affect the subsequent aging if the applied stress completely rejuvenates the material. Therefore any excessive or otherwise shear melting/history the material is subjected to, while loading the sample, is erased in the shear melting protocol. Cessation of shear melting step marks beginning of aging time ($t_w$=0). Similar to that observed for 2.8 weight % concentration system,[44] evolution of $G'$ typically follows a two-step evolution for other concentrations as well. Usually for experiments carried out on low idle times or with lesser salt concentrations, suspensions are observed to be in a liquid state ($G' < G''$) immediately after the shear melting is stopped. On the other hand, for experiments carried out on higher idle times or with greater concentration of salt, suspension directly enters a solid state after the shear melting ($G' > G''$). In the liquid regime $G'$ shows a stronger enhancement as a function of aging time compared to that of $G''$, and eventually crosses the same. The point of crossover ($G' = G''$) is represented as a liquid –solid transition in the literature.[32] It should be noted that this point of transition does depend on applied frequency (or a timescale of probe),[39] as also observed for glass transition in molecular glasses.[59] In solid state ($G' > G''$), $G'$ increases with a weaker dependence on aging time and $G''$ decreases after showing a maxima. We also perform aging experiments at different frequencies. Interestingly, $G'$ does not show any dependence on frequency beyond $G' > G''$. However $G''$ shows weak decrease with frequency in the solid state.

As mentioned before, evolution of $G'$ and $G''$ for experiments on Laponite suspension carried out at higher idle times and with greater concentration of salt shift to lower aging times. The representative behavior for a 3.5 weight % suspension is plotted in figure 2. It can be seen that the curvature of the evolution of $G'$ beyond the cross-over is self-similar, which leads to aging time – idle time superposition for various concentrations of salt as shown in figure 3(a). Various superpositions of $G'$ at different concentrations of salt also share self-similar curvature in the solid state ($G' > G''$) and produce aging time – idle time – salt concentration superposition for all the explored concentrations of Laponite. We report these superpositions in figure 4(a). Similar to the effect of idle time and salt concentration, evolution of elastic (and viscous) modulus at higher Laponite concentration shifts to lower aging times. In order to get aging time – idle time – salt concentration superposition at various Laponite concentrations shown in figure 4(a), we have primarily carried out the horizontal shifting of all the $G'$ evolutions. The vertical shift factor is required for salt concentration dependent shifting, though the value of $V_s$ is observed to be of the order of unity (vertical shift factor $V_L$ shown in figure 4(a) is merely used to show the superpositions for various Laponite concentrations clearly).

The horizontal shift factors ($h_i$) associated with idle time shifting (Shown in figure 3(b)) were observed to increase with increase in idle time. The salt concentration dependent horizontal shift factors ($H_S$) plotted as a function of concentration of salt for various concentrations of Laponite are shown in figure 4(b). It can be seen that the shift factors show an exponential dependence on salt concentration represented by: $\ln(H_S) \sim C_S$, irrespective of the Laponite concentration in the explored range of 2 to 3.5 weight %. It should be noted that various shift factors discussed in this work are not independent and separable from others and do depend on values of other variables. For example, in figure 3(a), $h_i = H(C_L, C_S, t_i, T)$, where $C_L = 3.5$ weight % and $T = 10°C$, and is plotted as a function of $C_S$ and $t_i$ in figure 3(b). Similarly in figure 3(a), $H_S h_t = H(C_L, C_S, t_{iR}, T_R)$, where $t_{iR} = 18$ day and $T_R = 10°C$ and is plotted as a function of $C_S$ for different $C_L$ for the specific values of $t_{iR}$ and $T_R$.

Although the evolution of $G'$ does demonstrate excellent superposition in the solid state (region of weak evolution of $G'$ at higher aging times, $G' > G''$), in the liquid state (region of strong evolution of $G'$ at lower aging times, $G' < G''$) the quality of superposition is poor. The primary reason for this could be inability of oscillatory flow experiments to access the evolution of viscoelastic behavior in the liquid region.[60] This point can be understood better by considering the rheological behavior of aqueous suspension of Laponite to be represented by a single mode Maxwell model (a spring and dashpot in series) with time dependent elasticity and viscosity. For a Maxwell model, $G'' = G'/(\omega\tau)$, where $\omega$ is frequency of oscillations and $\tau$ is characteristic (or dominating) relaxation time.[59] The relaxation time, which is very small in the liquid regime increases as a function of time (typically relaxation time shows exponential dependence on aging time in the liquid regime).[19, 53] As a result the point at which relaxation time becomes of the order of reciprocal of frequency ($\tau \approx 1/\omega$), $G'$ crosses over $G''$. Consequently in the liquid regime ($G' < G''$), relaxation time undergoes substantial increase over a period of one oscillation ($1/\omega$). Therefore oscillatory flow experiments tend to average the variation in viscoelastic properties over a duration of one cycle and thereby induce error while probing evolution of $G'$ and $G''$. Farther is the material in the liquid region from crossover ($G' = G''$), greater the error is.[60]

Similar to that of $G'$, evolution of $G''$ also shifts to lower aging times for experiments carried out for higher idle times and for higher concentrations of Laponite and salt. However, $G''$ shows a very peculiar behavior as a function of concentration of Laponite. In figure 5 we plot $G''$ for various concentrations of Laponite having 5mM salt and idle time of 6 days. It can be seen that slope with which $G''$ decreases on a double logarithmic scale (after the crossover $G' = G''$) becomes weaker with increase in concentration of Laponite. Such broadening of maxima for experiments carried out on greater idle times was reported recently.[53] The present observation of broadening with increase in Laponite concentration (over 2.4 to 3.5 weight %) is witnessed irrespective of concentration of salt present in the suspensions. On the other hand, $G''$ of 2 weight % suspension shows different behavior as shown in figure 6. For this concentration evolution of $G''$ demonstrates either a very weak decrease or a plateau after the crossover. However, unlike a prominent decrease in $G''$ observed for high concentrations of Laponite, at high aging time $G''$ in 2 weight % suspension can be seen to be increasing with the aging time.

We also study evolution of viscoelastic response at different temperatures for 2.8 weight % suspension with different concentrations of salt over a period up to 60 days after preparation of the same. In figure 7 (a & b) we plot evolution of $G'$ and $G''$ for 2.8 weight % Laponite suspension at different idle times and salt concentrations. Plot shown in figure 7(a) describes the evolution of $G'$ and $G''$ for a suspension without any externally added salt (0.1 mM) having idle time of 18 days. It can be seen that evolution of $G'$ and $G''$ shifts to lower aging times with increase in temperature. In figure 7(b), we plot aging behavior of a 9 days old 2.8 weight % 5 mM suspension. In this case suspension directly enters the glassy state ($G' > G''$) owing to faster aging dynamics in the presence higher amount of externally added salt. In figure 8, we have plotted temperature dependence of 2.8 weight % suspension at 5 salt concentrations for experiments carried out on day 12. On the other hand, in appendix figure S1 we have plotted evolution of temperature dependence as a function of idle time (day 6 to day 60) for a 2.8 weight % suspension having no salt (0.1 mM). As shown in figure 6 and figures 7 and 8, in the solid state, $G'$ follows the same trend of shifting to lower aging times at higher temperatures while preserving the curvature. However, evolution of $G''$ does not follow the same trend and shows weaker decrease for experiments carried out at higher temperatures. In figure 9 we plot $-d\ln G''/d\ln t_w$ for samples having 2.8 weight % Laponite and different concentrations of salt as a function of reciprocal of temperature, which clearly shows weaker slope at higher temperatures. In the solid state, self-similar curvatures associated with $G'$ suggest possibility of superposition upon horizontal shifting. In figure 7 (a & b) we represent this superposition by gray symbols. The corresponding shift factors $a_T$ are plotted as a function of reciprocal of temperature in figure 7(c).

The self-similar curvature of $G'$ in the solid state ($G' > G''$) irrespective of concentration of Laponite shown in figure 4(a) and that of as a function of temperature shown in figure 7 lead to *aging time – idle time – salt concentration – Laponite concentration – temperature* superposition. This superposition is plotted in figure 10 which contains 61 evolution curves obtained for different system variables. It should be noted that in order to get the superposition we need only horizontal shifting as a function of idle time, Laponite concentration, and temperature; while the vertical as well as horizontal shifting is required for salt concentration variation. In the inset of figure 10, we plot horizontal shift factor ($H_L$) as a function of Laponite concentration. Similar to that observed for salt concentration, horizontal shift factor associated with Laponite concentration shows an exponential dependence represented by $\ln(H_L) \sim C_L$.

We now discuss results of conductivity experiments. As discussed in introduction, particles of Laponite in an aqueous media share attractive as well as repulsive interactions among each other. As an apparent quantification of repulsive interaction, the electrostatic (Debye) screening length associated with the opposite faces of Laponite and its temporal evolution, if any, may help in understanding origin of irreversibility in this system. Debye screening length can be obtained with the knowledge of concentration of cations and anions present in the aqueous media. In order to estimate the same, we measure the conductivity of samples over a span of idle times up to 18 days as a function of concentration of Laponite and that of salt. For some samples we also measure conductivity as a function of temperature. The inset in figure 11 shows the ionic conductivity of 3.5 weight % Laponite suspension having various salt concentrations measured on different idle times. As expected, the conductivity of samples having greater amount of NaCl is higher. In the inset of figure 12 we plot conductivity as a function of concentration of Laponite having no externally added salt.[61] Conductivity can be seen to be increasing with increase in Laponite concentration owing to enhanced counter-ion concentration. However, very importantly, the conductivity can be seen to be continuously *increasing* with idle time over the explored range of 18 days in both the insets of figure 11 and 12. Since complexometric titrations have confirmed absence of $Mg^{2+}$ ions in the suspensions, continuous increase in ionic conductivity over 18 days may be attributed to slow dissociation of $Na^+$ counterions from Laponite particles. With knowledge of conductivity, the Debye screening length can be obtained by using an expression:[62]

$$\frac{1}{\kappa} = \left( \frac{\varepsilon_0 \varepsilon_r k_B T}{\sum_i (z_i e)^2 n_i} \right)^{1/2},$$ (1)

where $\varepsilon_0$ is permittivity of free space, $\varepsilon_r$ is relative permittivity, $k_B$ is Boltzmann constant, $z_i$ is the ionic charge number for $i$ th ion (NaCl is a 1:1 electrolyte), $e$ is the electron charge, and $n_i$ is the total number density of $i$ th ion. In order to evaluate the denominator we need to calculate the number density of Cl⁻ and Na⁺ ions. Since Cl⁻ ions are present in the suspension only through externally added NaCl, its concentration is same as that of NaCl. On the other hand, concentration of Na⁺ ions can be estimated from ionic conductivity and is given by: $\sigma = e\left(\mu_{Na}n_{Na} + \mu_{Cl}n_{Cl}\right)$,[63] where mobilities of Na⁺ and Cl⁻ ions are: $\mu_{Na}$ =5.19×10⁻⁸ m²/sV and $\mu_{Cl}$ =7.91×10⁻⁸ m²/sV.[64]

In figure 11 we plot the Debye screening length estimated for 3.5 weight % Laponite suspension as a function of salt concentration and idle time, while in figure 12 we plot the same as a function of concentration of Laponite and idle time for a no salt system. It can be seen that higher the concentration of salt, lower is the Debye screening length associated with a particle. Interestingly changing concentration of externally added Na⁺ ions from 0.1 to 7 mM causes decrease in Debye screening length by around 10 to 12 % on any studied idle time. Similarly the Debye screening length is observed to decrease with increase in concentration of Laponite. Finally and very importantly, Debye screening length is observed to be decreasing as a function of idle time over duration of 18 days in all the suspension samples due to continued dissociation of counterions. This proposal of continuous dissociation of counterions from Laponite particle, in addition to complexometric titration experiments, can also be supported by estimating maximum possible conductivity achievable for a suspension assuming all the counterions have been dissociated. Since a single particle of Laponite with 25 nm diameter contains around 1000 unit crystals and therefore the net negative charge on a Laponite particle is around 700 electronic charges.[3, 65] As a result, maximum possible Na⁺ ion concentration of a suspension, in addition to known amount of NaOH and NaCl added to the same, is 700 ions per particle. In figures 13 (a & b), we show that measured value of conductivity on day 18 is significantly smaller than the maximum possible conductivity. We also measure conductivity of suspension as a function of temperature by equilibrating the same with a temperature bath on a mentioned idle time. Interestingly, as shown in figure 14, conductivity can be seen to be increasing as a function of temperature. Consequently Debye screening length is expected to decrease at greater temperature.

## IV. Discussion:

Laponite particles in an aqueous medium share complex interactions among each other due to its anisotropic shape, dissimilar charges on the edge as well as the faces and screening of these charges due to presence of ions in the liquid media. These interactions may lead to repulsion between the faces and attraction between the edges and the faces of the particles. In addition, there could be van der Waals attraction among the particles. All these interactions and shape effects are responsible for microstructure of Laponite suspension, which evolves as a function of time causing changes in its viscoelastic character. According to various light scattering and rheological studies, in glassy (nonergodic) state relaxation time of Laponite suspension is known to have a power law dependence on aging time given by: $\tau = A\tau_m^{1-\mu}t_w^{\mu}$,[19, 27, 37, 38, 53, 66] where $\tau_m$ represents microscopic time scale associated structural reorganization (also represented as unit time with which structure builds). Power law dependence of relaxation time on aging time is known to be a generic feature of different types of glassy materials.[20, 60, 67-69] In Laponite suspension, dynamics of the system can be well represented by considering Laponite particles to be arrested in energy wells, typically represented as cages in the literature.[66] Under such situation, mere thermal energy is not sufficient for the particle to escape the cages.[60, 70] However, suspension can experience structural reorganization due to particles undergoing activated barrier hopping or by undergoing motion remaining inside the cage. Both these events take the suspension to progressively low energy state as a function of time. Assuming that all the particles occupy the energy wells having same depth $E$, relaxation time can be considered to have Arrhenius dependence on $E$ given by: $\tau = \tau_m \exp\left(E/kT\right)$.[60] The dependences of relaxation time on $E$ as well as aging time leads to: $E = kT \ln A + \mu kT \ln(t_w/\tau_m)$. If $b$ is characteristic length-scale associated with the suspension, a scaling relation for the elastic modulus of suspension was recently proposed by Shahin and Joshi[53] and is given by:

$$G' = \beta E/b^3 = \frac{\beta k_B T}{b^3} \ln A + \mu \frac{\beta k_B T}{b^3} \ln\left(t_w/\tau_m\right),$$  (2)

where $\beta$ is a proportionality constant. If we assume that viscoelastic behavior of suspension is given by single mode Maxwell model, viscous modulus of the same can be written as:[53]

$$\ln G'' = \ln G' - \ln\left(\omega A \tau_m\right) - \mu \ln\left(t_w/\tau_m\right).$$  (3)

Equation (2) suggests that evolution of $G'$ depends on aging time ($t_w$) normalized by microscopic time scale ($\tau_m$). $\tau_m$ can be assumed to have Arrhenius dependence given by: $\tau_m = \tau_{m0} \exp\left(U/k_B T\right)$, where $U$ is energy barrier related to microscopic movement within the cage (It should be noted that $E$ is depth of energy well associated with the cage while $U$ is activation barrier associated with microscopic timescale. In figure 15 we have represented this scenario using a schematic).

In the glassy domain, evolution of $G'$ can usually be described by, $G' = B + G_0 \ln\left(t_w/\tau_m\right)$ as shown by curvature of the superposition described in figure 10. Therefore, the power law index $\mu$ can be represented by:[53]

$$\mu = b^3 G_0 / \left(\beta k_B T\right),$$  (4)

where parameter $G_0$ does not depend on either concentration of Laponite or temperature as no vertical shifting is required with respect to both these variables to get the comprehensive superposition of $G'$ as shown in figure 10 (Vertical shift factor based on Laponite concentration used is figure 4 is just to show the respective superpositions more clearly. This shift factor has not been used in figure 10). Equation (3) qualitatively predicts that, in the glassy regime, $G''$ decreases with increasing aging time if the third term on the right side dominates the first term. According to this term $\left[-\mu \ln\left(t_w/\tau_m\right)\right]$, $\ln G''$ is expected to vary as a function of $\ln t_w$ with slope $-\mu$. According to equation 4, $\mu$ is proportional to $b^3$. If $b$ is considered to be average interparticle distance given by: $b = \left(\pi R^2 h / \phi_v\right)^{1/3}$, where $\phi_v$ is volume fraction of Laponite in suspension, $R$ is radius ($\approx 12$ to $15$ nm) and $h$ is thickness ($\approx 1$ nm) of Laponite particle,[44, 51] we get $\mu \sim G_0 / \left(\phi_v k_B T\right)$. This suggests that $\ln G''$ plotted as a function of $\ln t_w$ should decrease more steeply for lower concentrations of Laponite. Interestingly, as shown in figure 5, we indeed observe that $\ln G''$ decreases more steeply with decrease in concentration of Laponite. In addition, equations 3 and 4 also suggest that $\mu$ should decrease with increase in temperature. Remarkably we also observe this behavior as shown in figures 7 (b) and 9. For 2 weight % Laponite suspension, $G''$ is observed to be increasing as a function of aging time in the glassy domain as shown in figure 6. Since $G'$ is a continuously increasing function of aging time, increase in $G'' \left(= G' / (\omega \tau)\right)$ suggests that increase in $\tau$ is weaker than that of $G'$ for 2 weight % Laponite suspension.

In figure 10 we observe comprehensive superposition of elastic modulus irrespective of changes in studied variables. On the other hand, $G''$ does not show self-similar evolution as a function of Laponite concentration, temperature and idle time. However as discussed before, $G'$ can be directly related to energy well depth $E$ by considering the former to be energy density ($G' = \beta E / b^3$). Self-similarity in $G'$, therefore means self-similarity with which system explores low energy states thereby indicating microstructure buildup of aqueous suspension of Laponite is generic with respect to change in Laponite concentration, that of salt, aging time and temperature. The major difference with respect to all these variables is timescale of aging ($\tau_m$) related to these variables (time scale of aging is same as that associated with structural reorganization). Interestingly equation (2) qualitatively describes dependence of $G'$ on aging time ($t_w$) normalized by timescale of aging ($\tau_m$). The latter is assumed to show Arrhenius dependence [$\tau_m = \tau_{m0} \exp\left(U / k_B T\right)$]. In figure 10, in order to plot comprehensive superposition, aging time is multiplied by horizontal shift factors associated with Laponite concentration $(H_L)$, salt concentration $(H_S)$, idle time $(h_i)$ and temperature $(a_T)$. Since reference temperature is: $T_R$ =10°C, reference salt concentration is: $C_{SR}$ = 5mM, and reference idle time is $t_{iR}$ =18 day, we can represent the product of shift factors as: $H_L H_S h_i a_T = H(C_L, C_{SR}, t_{iR}, T_R)$. Overall, for arbitrary reference points, we can write: $\ln(\tau_m) \sim U/kT \sim -\ln\left[H(C_L, C_S, t_i, T)\right]$. In addition, it can be seen from figure 4(b) and the inset of figure 10 that $\ln[H_S] \sim C_S$ and $\ln[H_L] \sim C_L$ respectively. Consequently $U$ decreases linearly with concentration of Laponite and that of salt respectively in the explored range. Moreover, as shown in figure 3(b), $h_i$ is observed to increase with idle time, suggesting $U$ also to decrease with idle time. The effect of temperature is more complicated than the other three variables. Since increase in temperature increases conductivity of suspension as shown in figure 14, it directly affects energy barrier so that $U$ decreases with increasing temperature. On the other hand temperature also expedites the dynamics by thermal activation term in the assumed Arrhenius dependence. Overall, superposition of the rheological aging data described in figure 10 clearly shows that energy barrier associated with structural buildup decreases with idle time and temperature and decreases linearly with increase in concentration of Laponite and that of salt.

In addition to rheological study, measurement of conductivity and estimation of Debye screening length gives an independent approach to analyze interaction among Laponite particles as a function of studied variables. Figures 11, 12 and 14 show increase in ionic conductivity as a function of concentration of Laponite, that of salt, idle time and temperature. For these variables, except the salt concentration, enhanced conductivity is due to dissociation of Na$^+$ ions from the Laponite particles. Greater dissociation of Na$^+$ ions from the face of Laponite particle decreases Debye screening length as shown in figures 11 and 12. On the other hand such dissociation makes the face of Laponite particle more electronegative. Therefore in order to analyze effect of progressive dissociation of $Na^+$ ions from the faces on Laponite particles on the energetic interaction (attractive and repulsive) between the same, we solve DLVO theory for Laponite suspension. According to DLVO theory free energy per unit area is given by sum of double layer repulsion and van der Waals attraction.[62] The free energy per unit area between two layers of 2:1 clay (planner surfaces) is given by:[1]

$$W(d) = \left(\frac{64nk_BT}{\kappa}\right)\gamma^2 e^{-2\kappa d} - \frac{A_H}{48\pi}\left(\frac{1}{d^2} + \frac{1}{(d+\Delta)^2} - \frac{1}{(d+\frac{1}{2}\Delta)^2}\right)^2, \quad (5)$$

where $n$ are number of ions per unit volume, $d$ is half distance between the two plates, $A_H$ is Hamaker constant ($1.06\times10^{-20}$ J for Laponite[71]) and $\Delta$ is thickness of unit layers between the same plates (6.6 A).[1] Furthermore, $\gamma = \tanh\left(ze\Phi_0/4k_BT\right)$, where $z$ is valence of ion, and $\Phi_0$ is electric potential on the surface. In equation (5), the first term on the right hand side represents double layer repulsion while the second term represents van der Waals attraction between the two layers.

In order to compute free energy using equation (5), we assume complete delamination of all the Laponite particles. The number of sodium ions dissociated per particle is given by: $(n_{Na} - n_0)/n_P$, where $n_0$ is concentration of $Na^+$ ions due to external sources (externally added NaOH or NaCl) and $n_P$ is number of particles per unit volume. In equation (5) number of ions per unit volume are therefore related to electric potential on the surface ($\Phi_0$) by: $\Phi_0 = e(n-n_0)/(2A_Ln_P)$, $A_L$ is area of one face of Laponite particle. In figure 16 we plot the total free energy per unit area along with respective contribution from double layer and van der Waals interaction for 3.5 weight % Laponite suspension without externally added salt for idle times: 3, 12 and 18 day. It can be seen that van der Waals attraction is significantly weaker compared to the double layer repulsion except when particles are very close to each other. Interestingly with increase in idle time, owing to higher value of coefficient $\gamma^2$ arising from greater negative surface potential ($\Phi_0$), height of repulsive barrier that particles need to cross in order to approach each other in parallel fashion increases. On the other hand, width of the repulsive barrier becomes narrower as a function of idle time due to stronger exponential decay ($e^{-2\kappa d}$) caused by decrease in Debye screening length ($1/\kappa$). We expect qualitatively similar behavior as shown in figure 16 when idle time is replaced by temperature. For the case of externally added salt ($C_S$), increase in $C_S$, for the same extent of dissociation of counterions, will decrease Debye screening length further without affecting value of coefficient $\gamma^2$. However since, $\kappa \sim n^{0.5}$ the coefficient of double layer repulsion term ($nk_BT/\kappa$) given by equation (5), will still scale as $nk_BT/\kappa \sim n^{0.5}$. This would, therefore, lead to increase in height of the repulsive barrier while making its width narrower with increase in $C_S$ when particles approach each other in a parallel fashion. Therefore the scenario will be qualitatively similar to that described by figure 16 when increase in idle time is replaced by increase in $T$ as well as $C_S$.

The results of rheological study suggest decrease in energy barrier for structure formation with increase in idle time, salt concentration and temperature (timescale associated with increase in $G'$ shifts to lower times with increase in $t_i$, $C_S$ and $T$). Therefore, although height of repulsive energy barrier is increasing, the fact that width of the same is shrinking (reduction in Debye screening length) with increase in $t_i$, $C_S$ and $T$ is intuitively in disagreement with the proposal of low energy structures being repulsion dominated. In order to quantitatively probe effect of repulsive interactions on the low energy structure we analyze the timescale associated with structure formation at which material undergoes liquid – solid transition with respect to Debye screening length. In figure 18, we plot aging time at which $G'$ crosses over $G''$ as a function of normalized Debye screening length (Debye screening length divided by average interparticle distance ($1/(b\kappa)$) for 1 mM and 3 mM salt concentration suspensions having different Laponite concentrations (For 2.4, 2.8, 3.2 and 3.5 weight % suspension, value of $b$ is approximately 42, 40, 39, 37 nm respectively). We represent this crossover time by $t_w^*$. As mentioned before, if we assume material behavior to be represented by a time dependent single mode Maxwell model, the point at which $G'$ crosses over $G''$ corresponds to $\tau = 1/\omega$. Since, in all the experiments we have employed fixed $\omega$ =0.1 Hz, $t_w^*$ represents the time at which dominating relaxation time of the suspension reaches a level of 10 s. It can be seen in figure 16 that the Debye screening length is around or less than 10 % of that of average inter-particle distance. For a given concentration of Laponite and salt, points from right to left suggest increase in idle time, which causes decreases in Debye screening length (or increase in electronegative surface potential) as shown in figures 11 and 12. Figure 17 shows that $t_w^*$ continuously decreases with decrease in Debye screening length. Furthermore, for a given concentration of Laponite, increase in $C_S$ shifts the dependence of $t_w^*$ on $1/(b\kappa)$ to lower values of $t_w^*$. If low energy structures are repulsive in origin we would have expected $t_w^*$ to be a decreasing function of $1/(b\kappa)$, contrary to experimental observation. In addition, overall behavior described in figure 17 suggests that concentration of Laponite and that of salt affect the dynamics independently in addition to Debye screening length. In purely repulsive interactions we would have expected $t_w^*$ to demonstrate a superposition as a function of $1/(b\kappa)$ irrespective of various system variables.

While undergoing thermal motion, rather than approaching in a parallel fashion, if two Laponite particles approach each other in a perpendicular fashion or from sides as shown in schematic figure 17, owing to dissimilar charges on the edge and on the face, we intuitively expect repulsive energy barrier to reduce with increase in negative surface potential ($\Phi_0$) and/or reduction in Debye screening length. Interestingly, this proposal agrees well with the findings of the rheological study. When particles approach each other in a perpendicular fashion, they form house of cards structure while when they approach each other from sides they form of overlapping coin configuration. Interestingly, a very recent Monte Carlo simulation study inspired by Laponite suspension report mixture of both these configurations.[16] Although simply based on rheological and conductivity results it is difficult to comment on precise microstructure, various results discussed in this paper make it amply clear that low energy structures associated with aqueous suspension of Laponite are influenced by attraction among Laponite particles. In the literature it is clearly established that addition of monovalent salt such as NaCl in aqueous suspension of Laponite increases dominance of attraction.[42, 43, 51] In this paper we extend this proposal and state that increase in Laponite concentration, temperature and very importantly the time elapsed since preparation of Laponite suspension (idle time) also lead to the microstructures that are influenced by attraction.

In the discussion so far we assumed complete delamination of Laponite particles in short time followed by slow but progressive dissociation of counterions. This was considered to be a cause for enhancement in ionic conductivity as a function of time. However, such increase in conductivity could also be attributed to slow delamination of Laponite particles (delamination to be rate determining step), in which the dissociation of counterions can be considered to be faster. In this case electronegative surface potential ($\Phi_0$) can be assumed to remain constant throughout the process. However, dissociation of counterions would certainly increase $n$ thereby leading to qualitatively same free energy scenario as shown in figure 16. The important difference would be increase in effective particle density as a function of time, which through attractive interactions could lead to enhanced shear modulus.

Shahin and Joshi[44] had reported partly irreversible nature of aging in aqueous suspension of Laponite. We feel that physicochemical effects such as irreversible increase in ionic conductivity as a function of idle time, which progressively reduces width of repulsive barrier (thereby enhancing rate at which attractive interactions may form) among the Laponite particles, and inability of high shear to rejuvenate the evolved microstructure are responsible for irreversible aging in aqueous suspension of Laponite. Recently Shahin and coworkers observed very strong anisotropic orientation near the air – Laponite suspension interface by observing birefringence when kept in between crossed polarizers.[52] The anisotropic orientation and its penetration into the bulk were observed to be getting pronounced as a function of concentration of Laponite, concentration of salt, temperature, and idle time. Interestingly, the same variables are responsible for faster aging in this system. We therefore believe that anisotropic orientation near the interface, its penetration in the bulk are related to enhanced conductivity (or reduced Debye screening length or repulsion) in the suspension. In addition, we feel that results discussed in this paper are in agreement with the experiments of Ruzicka and coworkers[72] who observed dissolution of young (around 2 to 3 days) Laponite suspension having 3 weight % concentration (upon addition of deionized water).[73] However, for older samples (~ 7 day) suspension didn't dissolve but underwent swelling when deionized water was added on top.[72] They proposed that repulsion is prevalent among the particles in the initial period while attraction develops in the Laponite suspension over long durations. The present study provides a plausible explanation to this observation.

**V. Conclusion:**

In this work we carry out an extensive study of aging behavior of aqueous suspension of Laponite using rheological and conductivity measurements, as a function of concentration of Laponite (2 to 3.5 weight %), concentration of salt (NaCl, 0.1 to 7 mM), and temperature (10 to 40°C) at regular intervals up to 60 days after preparation of the suspension (represented as idle time). We observe that rheological experiments carried out on greater idle times do not rejuvenate the suspension to the same initial state upon application of strong shear suggesting irreversibility in structural build-up. Furthermore, evolution of elastic and viscous modulus (and their cross-over), subsequent to shear melting, shift to smaller aging time for experiments carried out on greater idle times, higher concentrations of salt as well as that of Laponite, and temperature. Self-similar evolution of elastic modulus, which shows monotonic increase with aging time, leads to *aging time – idle time – salt concentration – Laponite concentration – temperature superposition* in the solid regime. Existence of such comprehensive superposition suggests generic nature of the microstructure buildup upon change in above mentioned variables over the explored range. On the other hand, in the solid regime viscous modulus is observed to decrease as a function of aging time with more pronounced slope for experiments carried out at lower temperatures and on suspensions having lower concentration of Laponite. Our simple scaling model qualitatively explains this behavior. Variation of shift factors necessary to get superposition of elastic modulus suggests that energy barrier associated with structure formation goes on decreasing with temperature and idle time, and particularly decreases linearly with increase in concentration of Laponite and that of salt. Interestingly, the conductivity experiments carried out on suspensions show continuous increase in conductivity and therefore continuous decrease in Debye screening length as a function of idle time. Analysis of interparticle interaction using DLVO theory suggests that when particles approach each other in a parallel fashion, height of repulsive energy barrier increases while narrowing its width with increasing idle time, temperature and salt concentration. However since edge and face of Laponite particle has dissimilar charges, attraction between the same is expected to increase when particles approach each other in a perpendicular fashion. Analysis of rheological and conductivity data, therefore strongly indicates influence of attractive interactions in forming the low energy structures in aqueous suspension of Laponite.

**Acknowledgement:** Financial support from Department of Science Technology, Government of India through IRHPA scheme is greatly acknowledged.

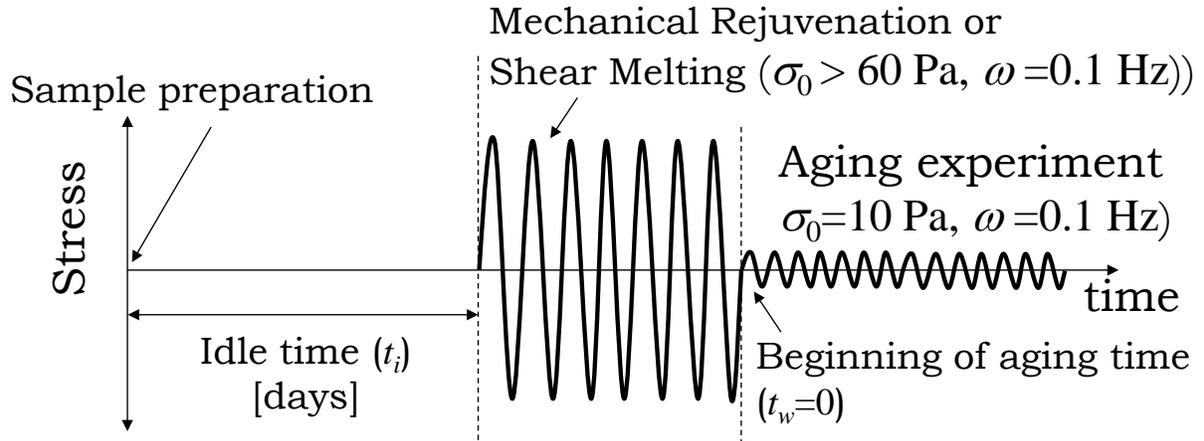

**Figure 1.** Schematic representing experimental protocol employed for the rheological experiments.

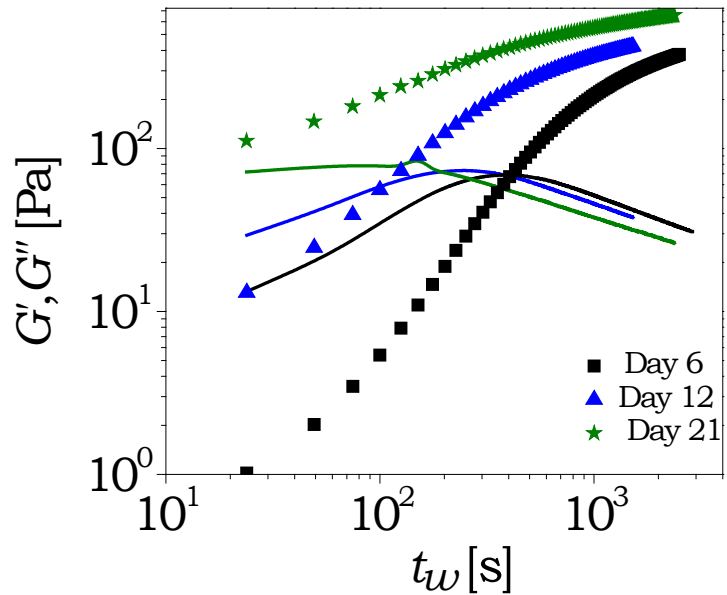

**Figure 2.** Evolution of Storage (symbols) and loss modulus (lines) on different idle times ($t_i$) for a suspension having 1mM salt and 3.5 weight % Laponite.

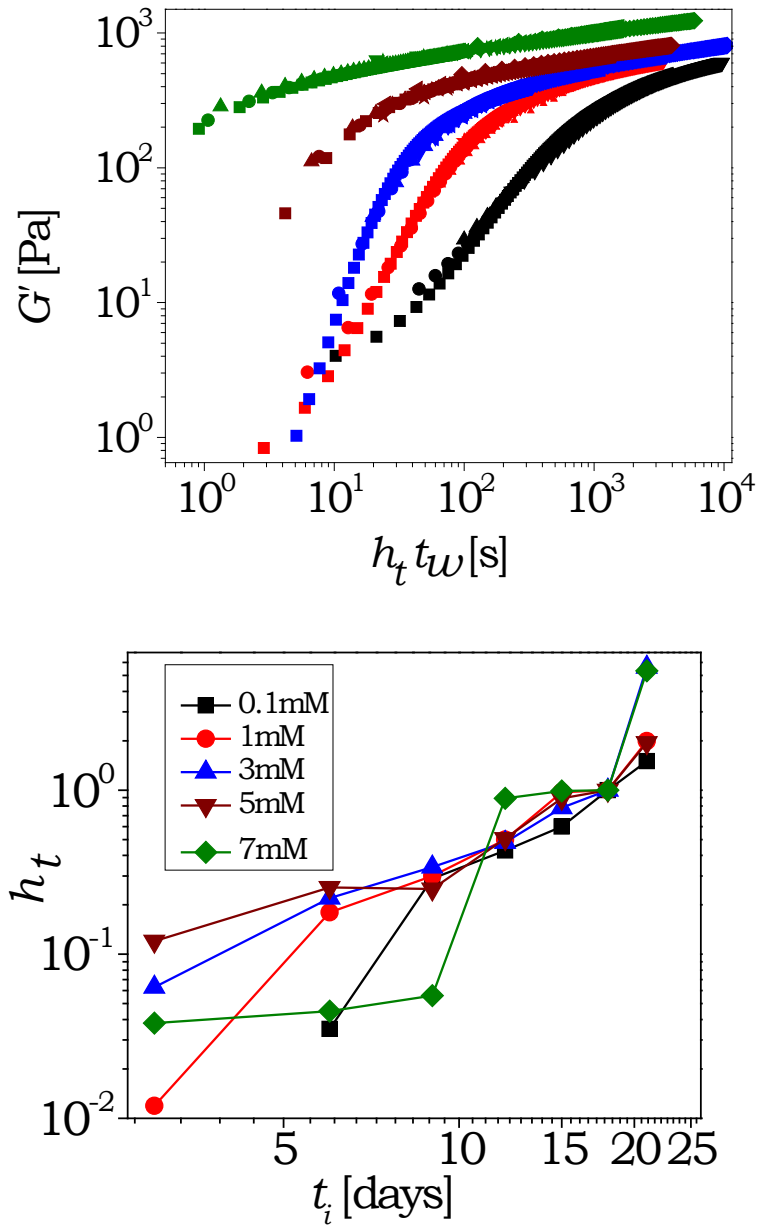

**Figure 3 (a).** Time – aging time superpositions of storage modulus for various salt concentrations having 3.5 weight % Laponite. From right to left: 0.1, 1, 3, 5 and 7mM salt concentration. Evolutions on various idle times are horizontally shifted on $t_i$ = 18 day evolution to obtain superpositions for different salt concentrations. (**b**) Horizontal shift factor required to obtain superposition is plotted as a function of idle time ($t_i$) for storage modulus data shown in figure 2a.

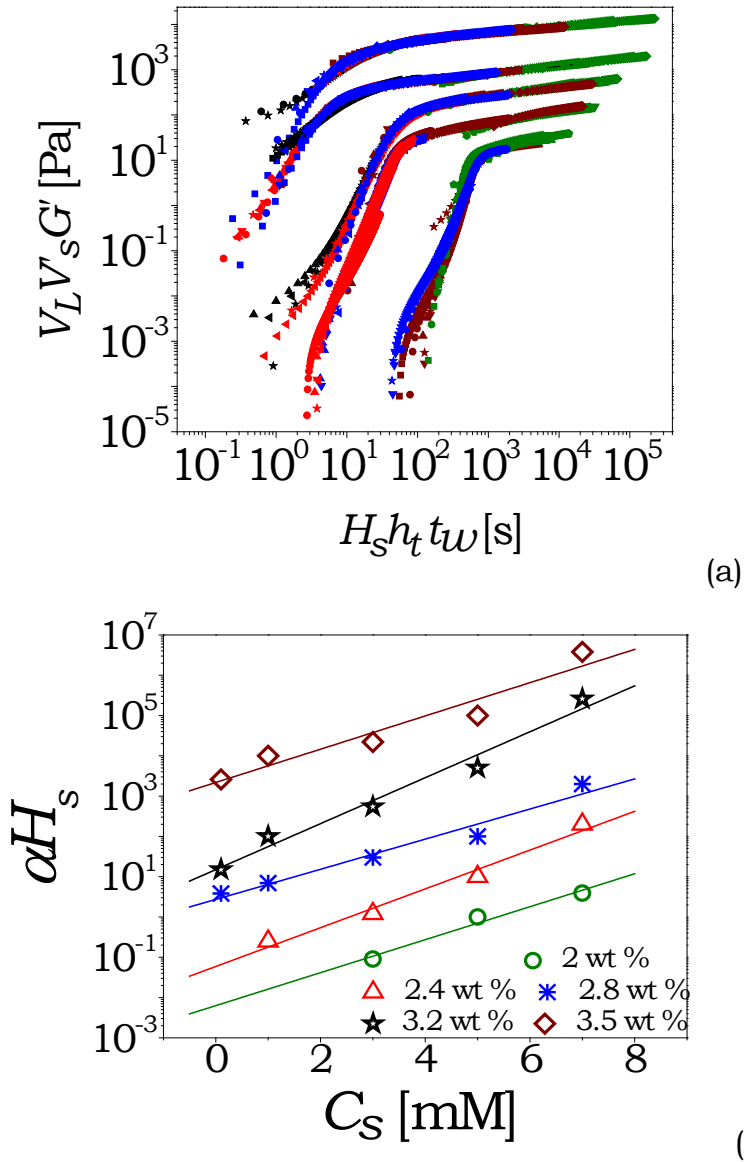

(a)

(b)

**Figure 4 (a).** Aging time – idle time – salt concentration superposition for evolution of $G'$ for various Laponite concentrations. From right to left: 2, 2.4, 2.8, 3.2 and 3.5 weight %. The curves have been shifted vertically with respect to concentration of Laponite for better clarity. **(b)** Dependence of horizontal shift factor based on salt concentrations used for obtaining the superposition shown in (a). Line fitted to the data represents $\ln(H_s) \propto C_s$. The shift factor data has been shifted vertically with $\alpha$ =1, 10, $10^2$, $5\times10^3$, $10^5$ for 2, 2.4, 2.8, 3.2 and 3.5 weight % respectively for better clarity.

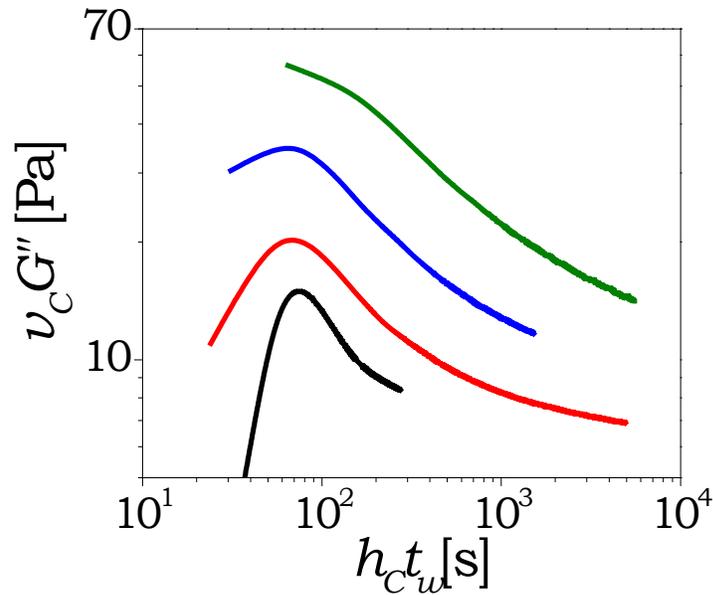

**Figure 5.** Evolution of Loss modulus for a suspension having 5 mM salt and an idle time of 6 days with varying Laponite concentrations. From top to bottom: $C_L$ = 3.5, 3.2, 2.8, and 2.4 weight %. The respective concentration curves have been shifted in both horizontal and vertical directions to describe the phenomenon more clearly.

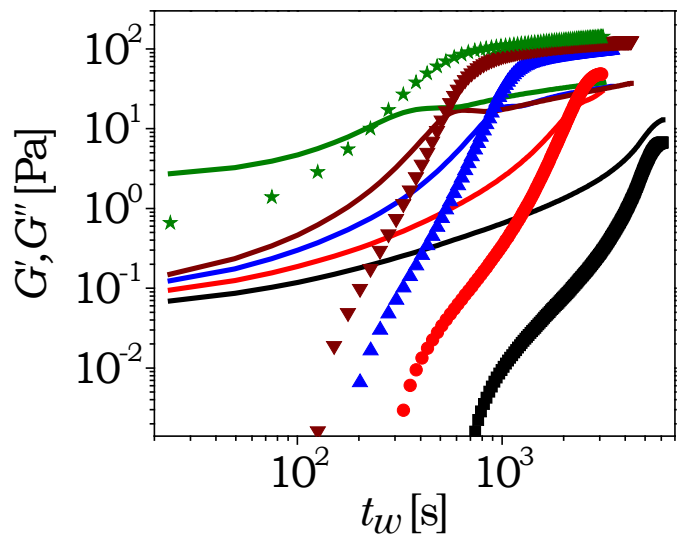

**Figure 6.** Evolution of $G'$ (filled symbols) and $G''$ (Lines) is plotted as a function of aging time on various idle times ($t_i$) for a suspension having 2 weight % Laponite and 5mM salt. From right to left: $t_i$ = 6, 9, 12, 15, 18 days.

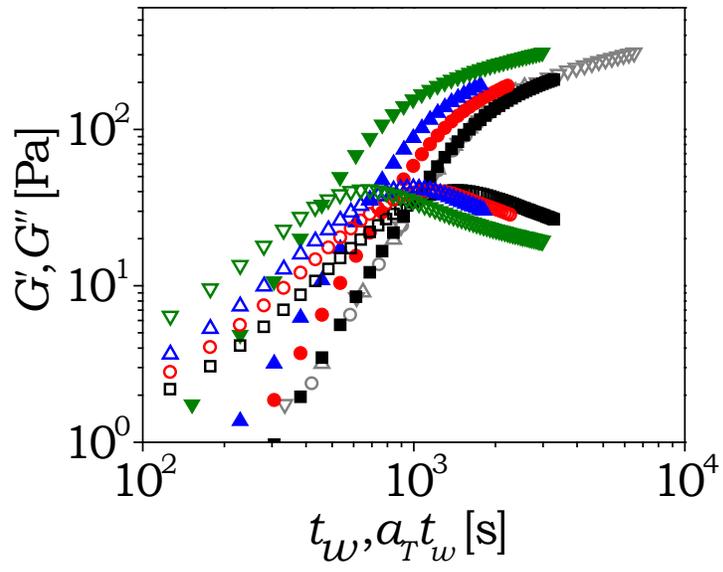

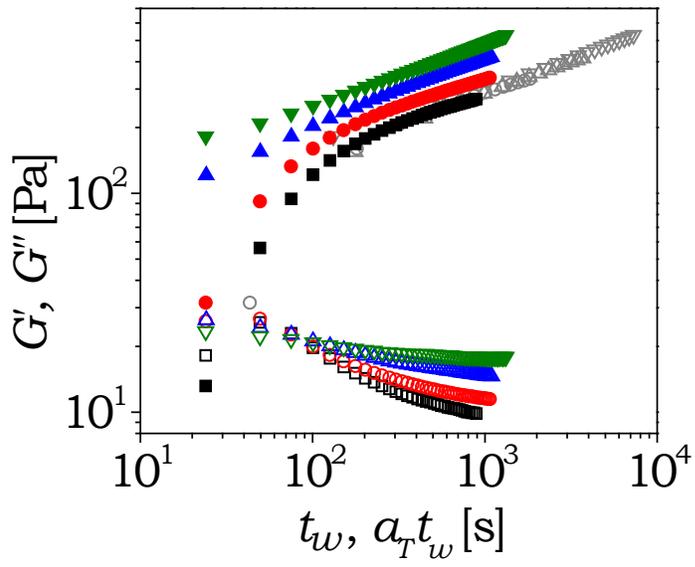

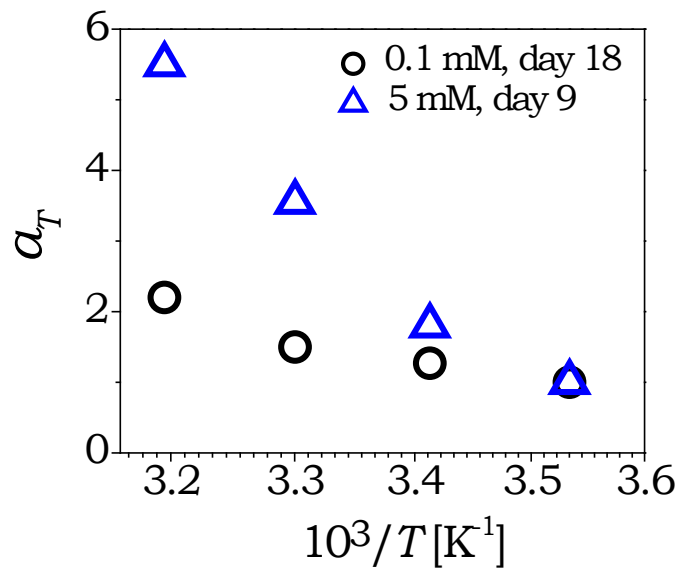

(c)

**Figure 7.** Evolution of $G'$ (filled symbols) and $G''$ (open symbols) is plotted as a function of aging time for different temperatures for (a) no salt system having 2.8 weight % Laponite and idle time of 18 days. (0.1mM); and (b) 2.8 weight % Laponite and 5mM salt and idle time of 9 days (Black 10 °C, Red 20 °C, Blue 30 °C and Green 40 °C). The open grey symbols show aging time – temperature superposition of $G'$. The corresponding shift factors are plotted with respect to reciprocal of temperature in (c).

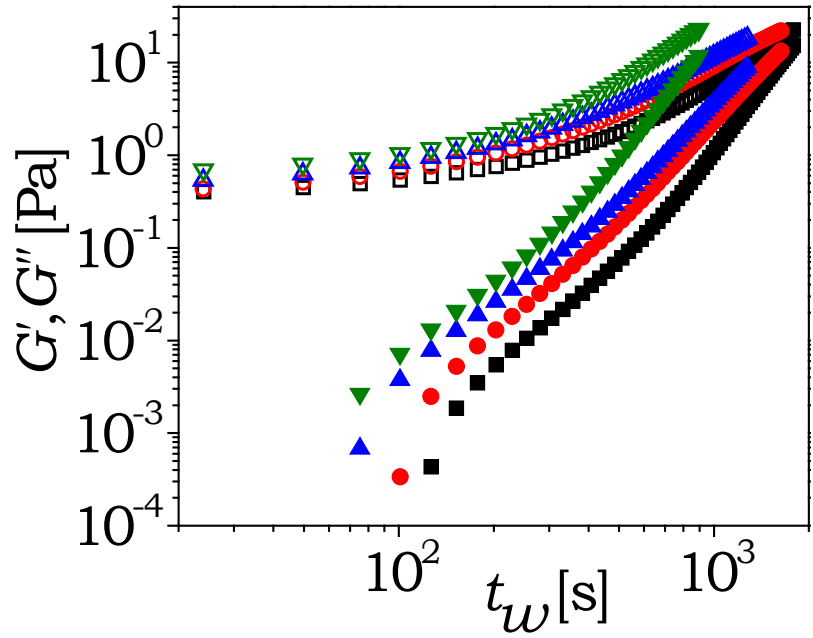

(a)

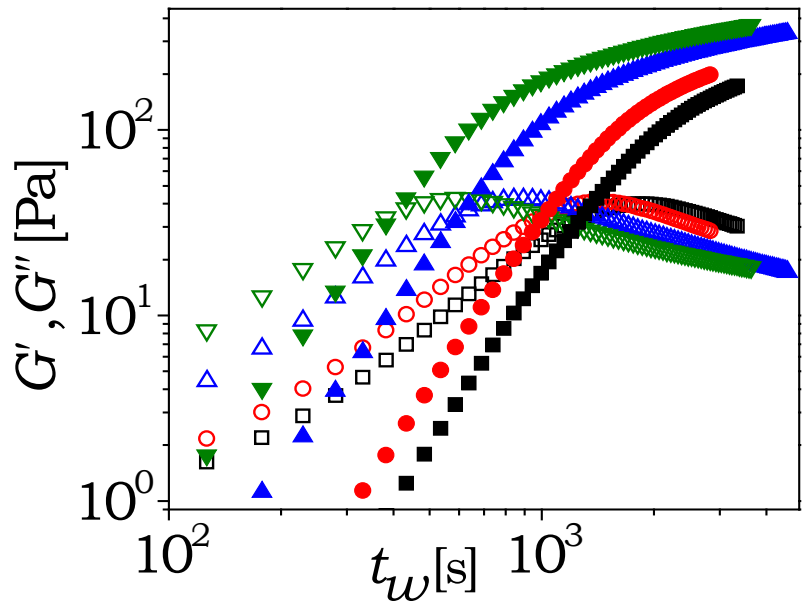

(b)

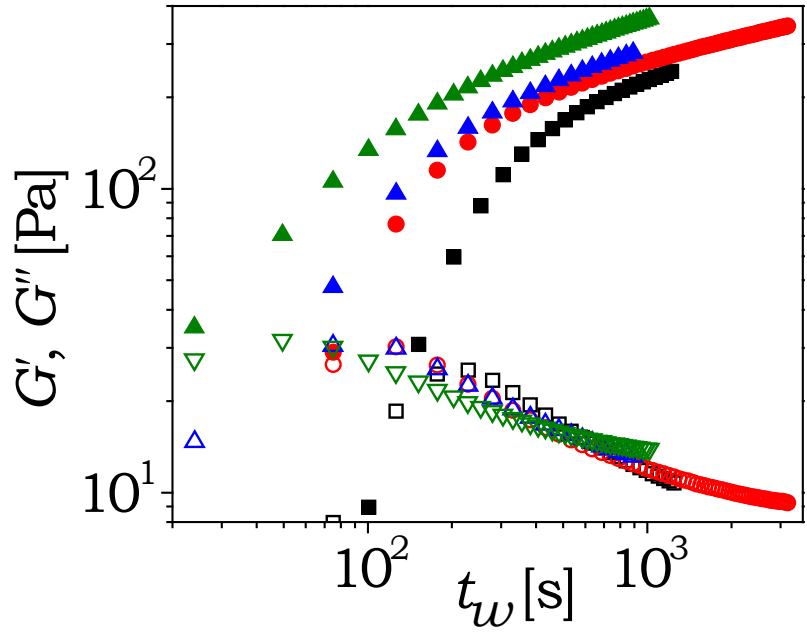

(c)

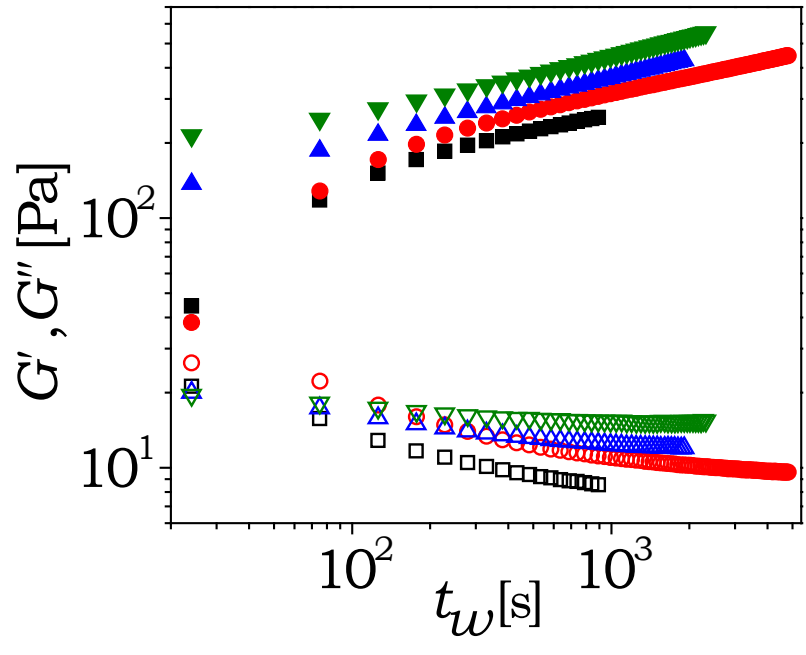

(d)

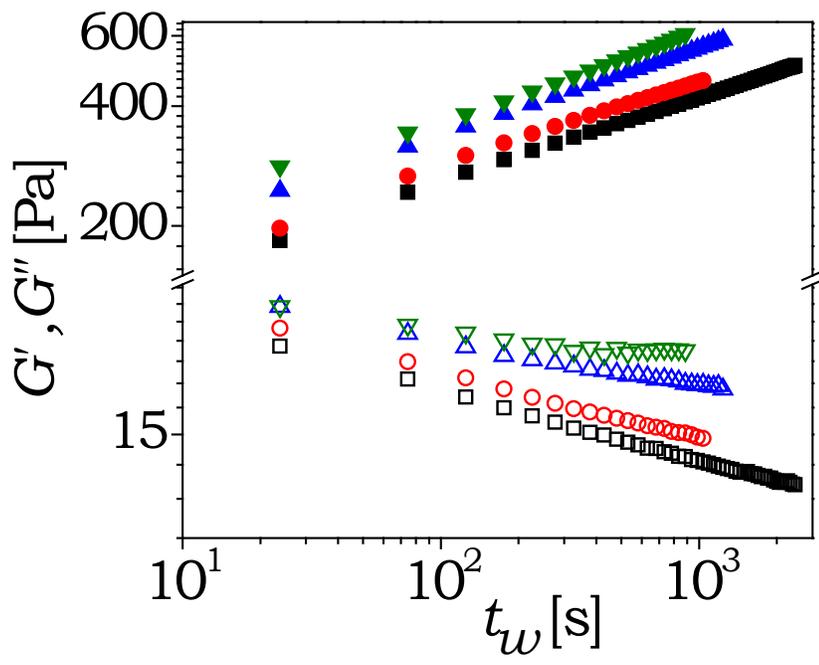

(e)

**Figure 8.** Evolution of storage (closed symbols) and Loss (open symbols) modulus at various temperatures for a system having 2.8 weight % Laponite and $t_i$=12 days. (a) 0.1 mM (b) 1 mM (c) 3 mM (d) 5 mM (e) 7 mM salt (Black squares 10°C, Red circles 20°C, Blue up triangles: 30°C and Green down triangles 40°C).

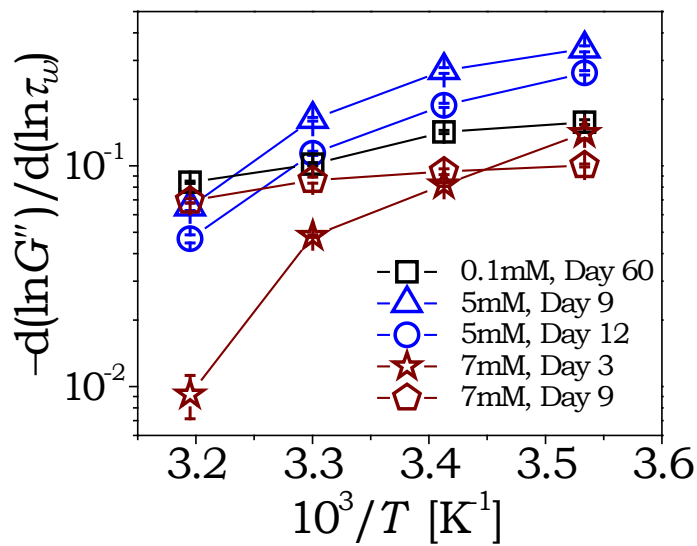

**Figure 9.** Variation of the slope of the loss modulus with increase in temperature for various idle times and salt concentrations.

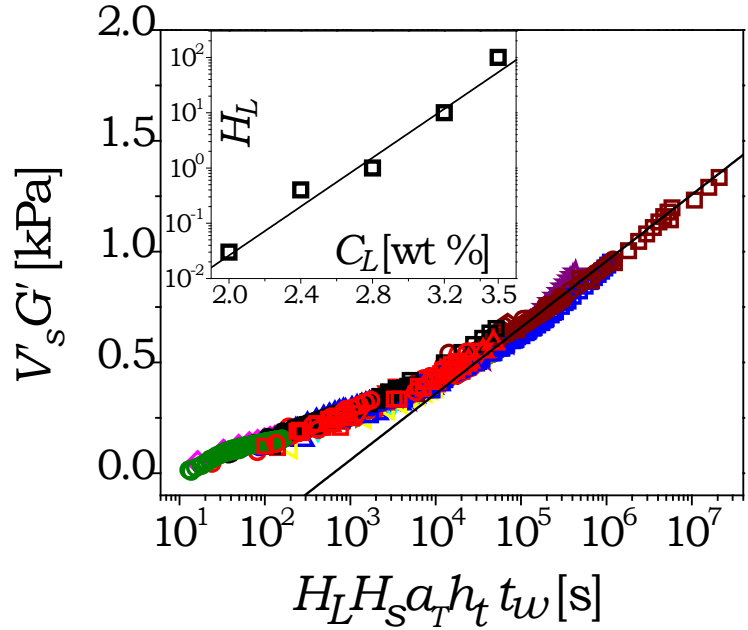

**Figure 10.** Aging time – idle time – salt concentration – Laponite concentration – temperature superposition for $G'$. The superposition contains 61 curves obtained at different idle times, Laponite concentrations, salt concentrations and temperatures. Only the data beyond the crossover of $G'$ over $G''$ is considered for the superposition. Black line passing through superposition represents $G' \sim \ln(t_w)$. Inset shows horizontal shift factors $H_L$ used to obtain the superposition as a function of concentration of Laponite. The straight line through the data plotted on semi-logarithmic scale demonstrates: $\ln(H_L) \sim C_L$.

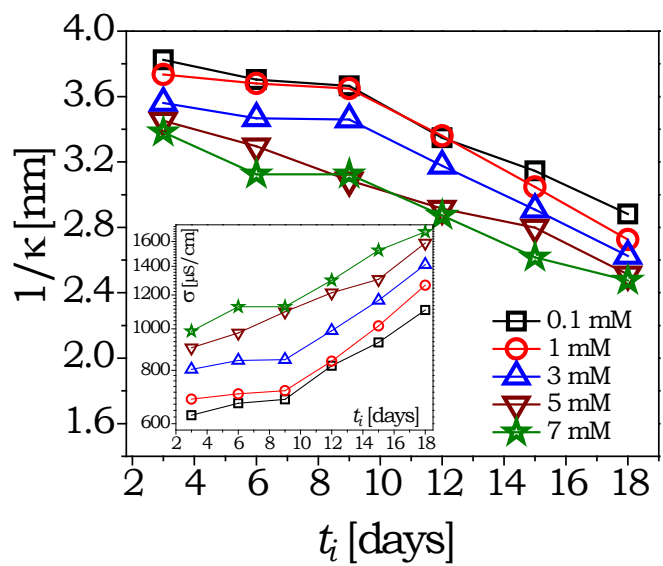

**Figure 11.** Debye screening length for 3.5 weight % Laponite suspension having different concentrations of salt as a function of idle time. Inset shows the ionic conductivity for the same.

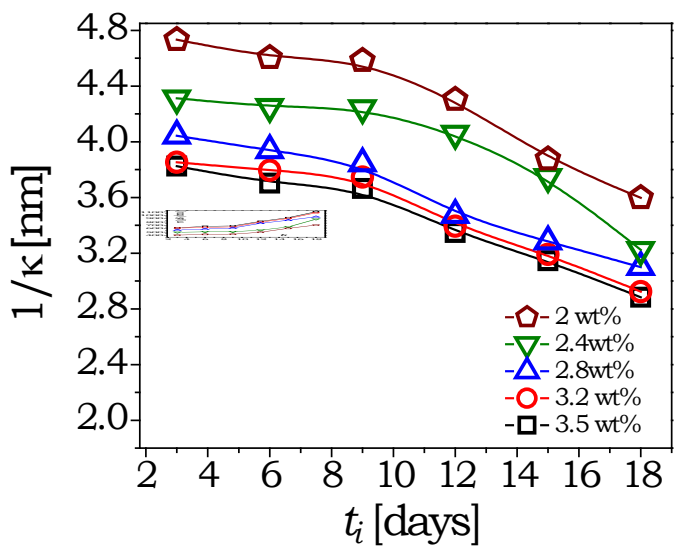

**Figure 12.** Debye screening length for various Laponite concentrations having no salt as a function of idle time. Inset shows ionic conductivity for the same.

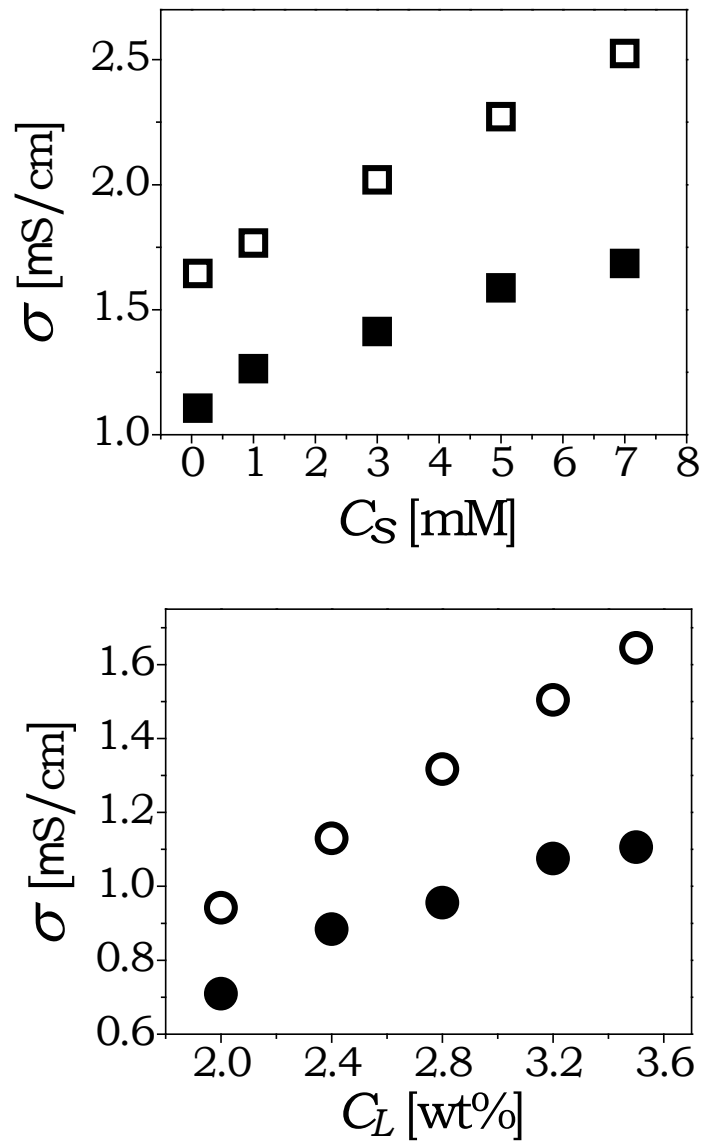

**Figure 13.** Ionic conductivity is plotted as a function of (a) salt concentration for 3.5 weight % Laponite suspension and (b) concentration of Laponite for no externally added salt (0.1 mM). Closed symbols represent measured ionic conductivity on an idle time of $t_i$=18 days while open symbols represent theoretically calculated maximum possible conductivity.

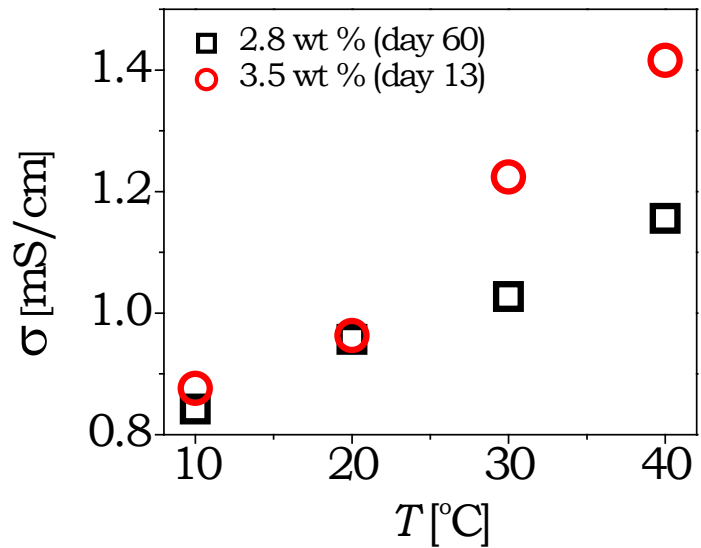

**Figure 14.** Ionic conductivity of aqueous suspension of Laponite as a function of temperature for a 3.5 weight %, 0.1 mM suspension on day 13, and for 2.8 weight % 0.1 mM suspension on day 60.

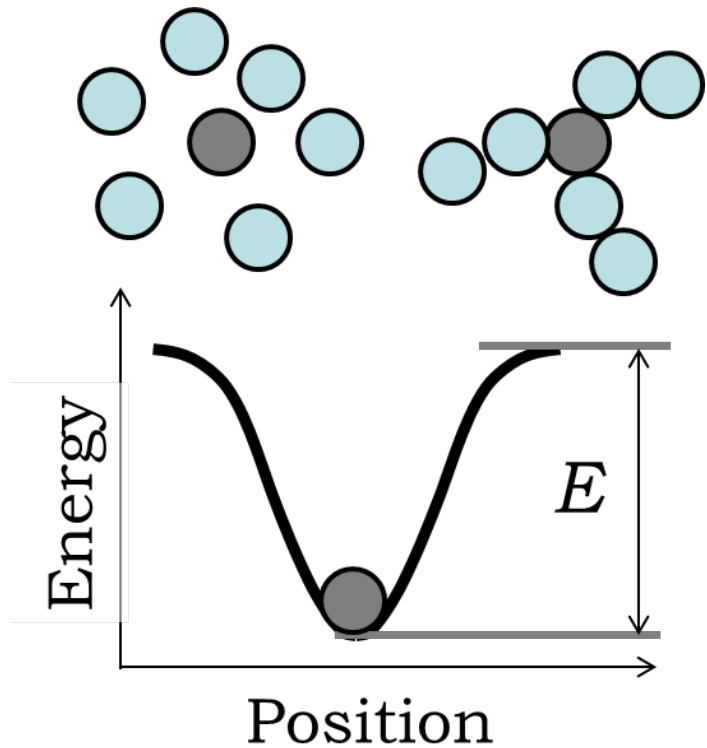

**Figure 15.** A schematic representing a particle under consideration arrested in a cage-like environment. Irrespective of glass–like or gel–like microstructure cage can be represented as an energy well having depth $E$. Particle undergoes microscopic motion of structural rearrangement within the cage with time scale $\tau_m$ such that $E$ goes on increasing as a function of time. $\tau_m$ is assumed to have Arrhenius dependence on temperature: $\tau_m = \tau_{m0}\exp\left(U/k_B T\right)$, where $U$ is energy barrier associated with structure formation. Since $\tau_m$ sets the time scale of enhancement of energy depth $E$, it evolves as: $E = E\left(t_w/\tau_m\right)$.

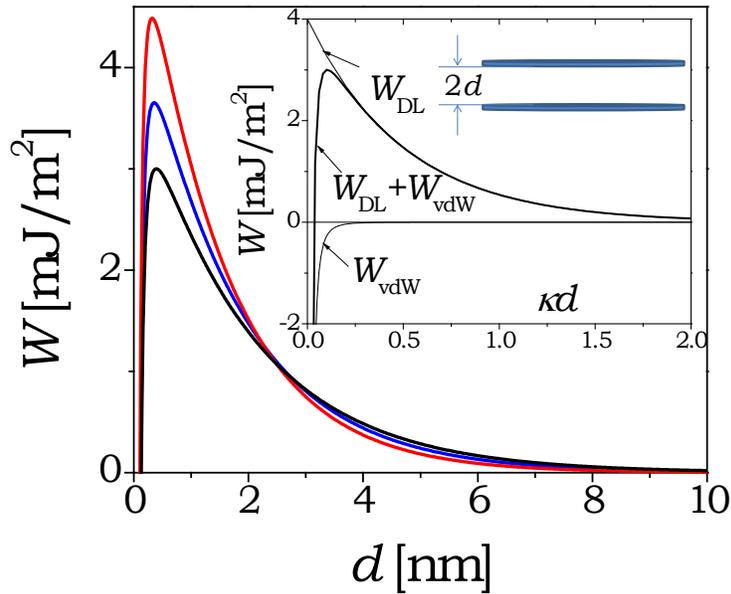

**Figure 16.** Free energy per unit area between two layers of 2:1 clay (planar surfaces) from DLVO theory is plotted as a function of half distance between the clay layers for 3.5 weight % Laponite suspension (without any externally added salt) on different idle times: Black line (smallest peak and weakest decay, $\sigma$ =628 µS/cm and $1/\kappa$ =3.8 nm) represents day 3, Blue line (middle line, $\sigma$ =820 µS/cm and $1/\kappa$ =3.3 nm) represents day 12, while Red line (highest peak and fastest decay, $\sigma$ =1106 µS/cm and $1/\kappa$ =2.9 nm) represents day 18. The inset shows variation of free energy per unit area for double layer interaction, van der Waals interaction and their sum as a function of half distance between the clay layers normalized by Debye screening length for 3.5 weight % Laponite suspension (without any externally added salt) on day 3.

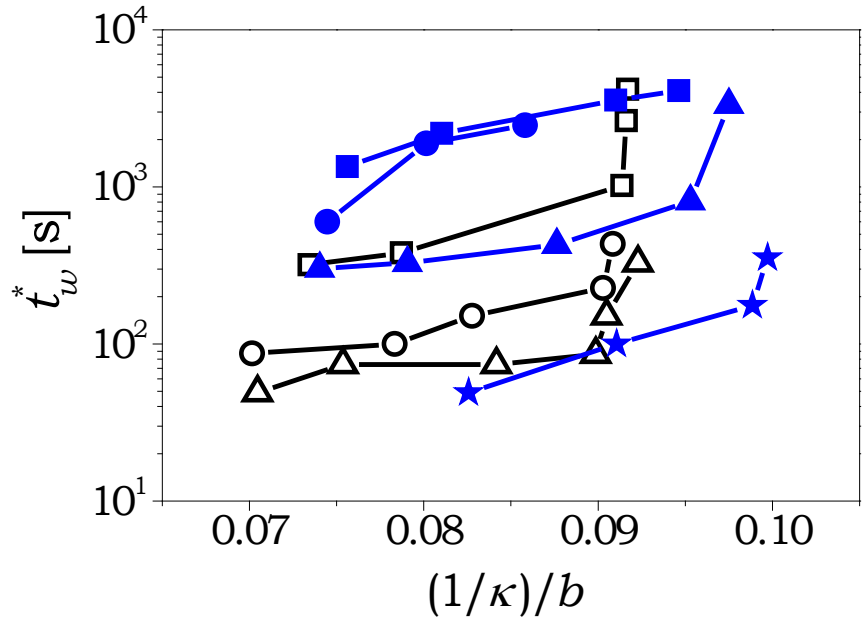

**Figure 17**: Aging time at cross over ($t_w^*$) of $G'$ and $G''$ is plotted against normalized Debye screening length for various Laponite concentrations [Square: 2.4 wt. %, Circles: 2.8 wt. %, Triangles: 3.2 wt. % and Stars: 3.5 wt. %]. Filled blue symbols represent 1 mM NaCl concentration while open black symbols represent 3 mM concentration.

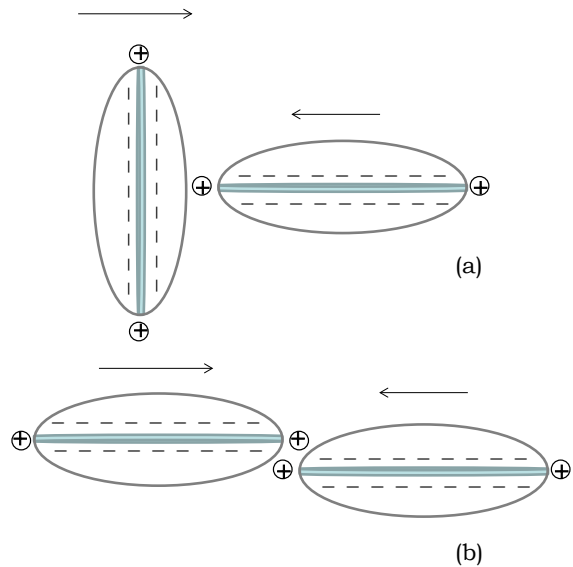

**Figure 18.** Schematic of a scenario when two Laponite particles, owing to thermal motion, approach each other (a) in a perpendicular fashion and (b) from the sides.

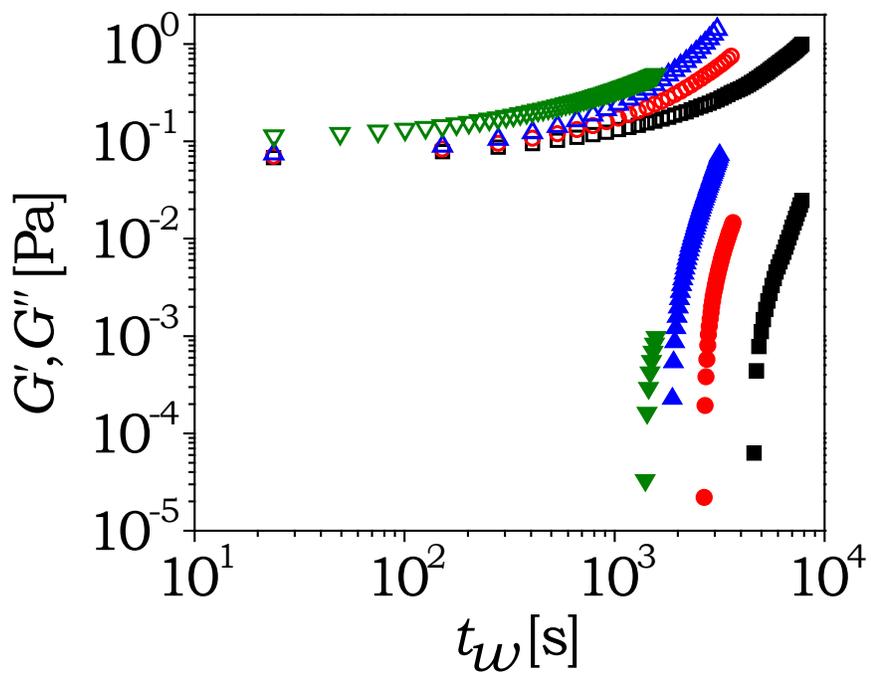

(a)

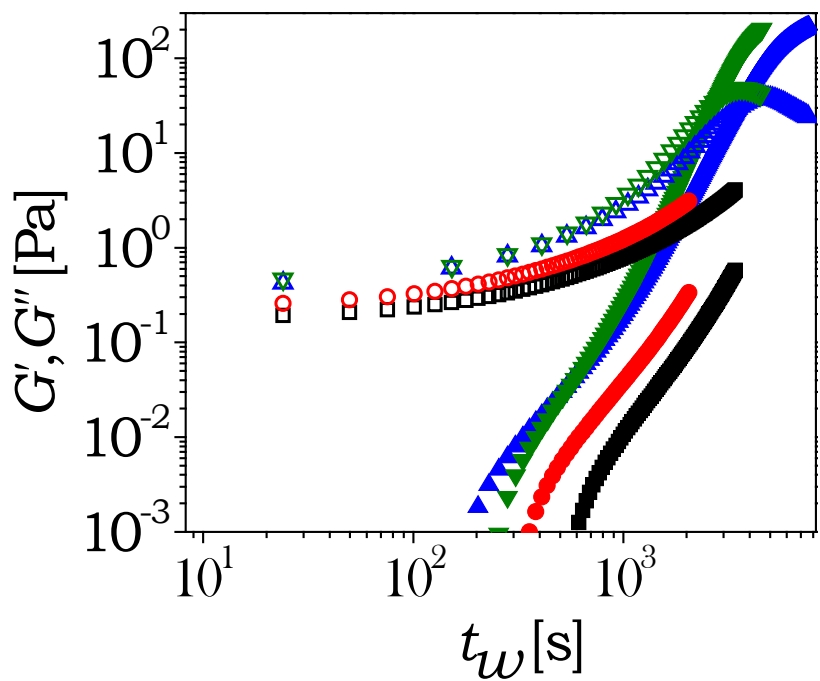

(b)

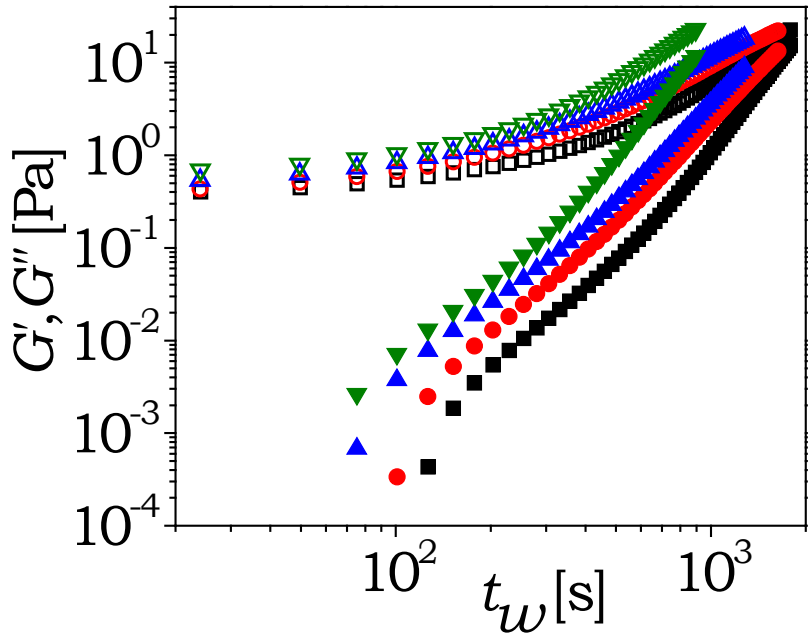

(c)

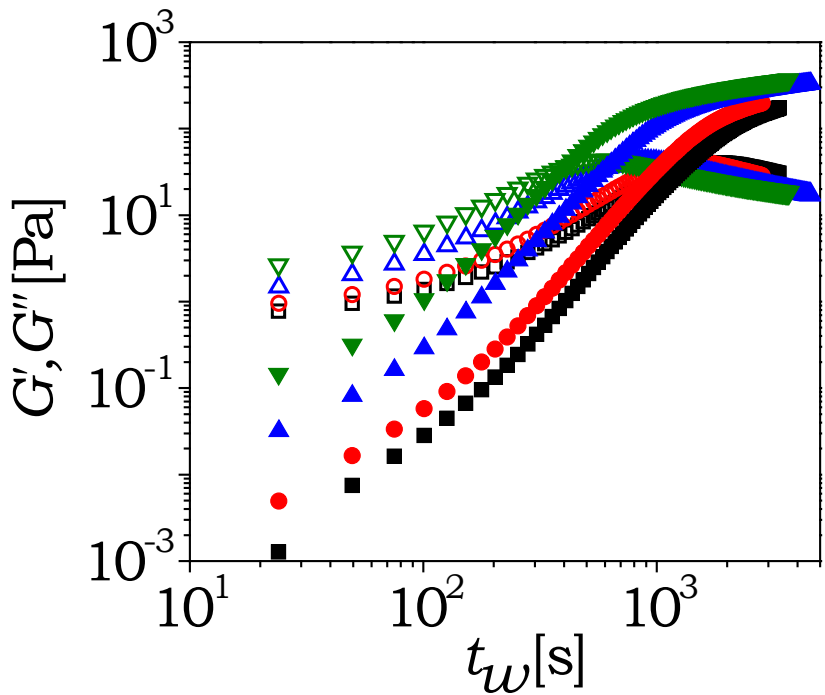

(d)

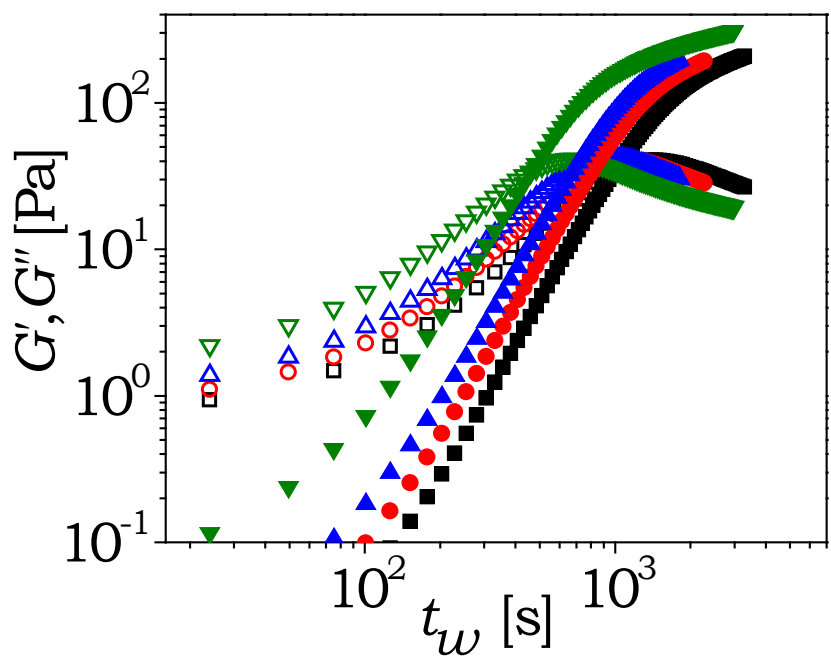

(e)

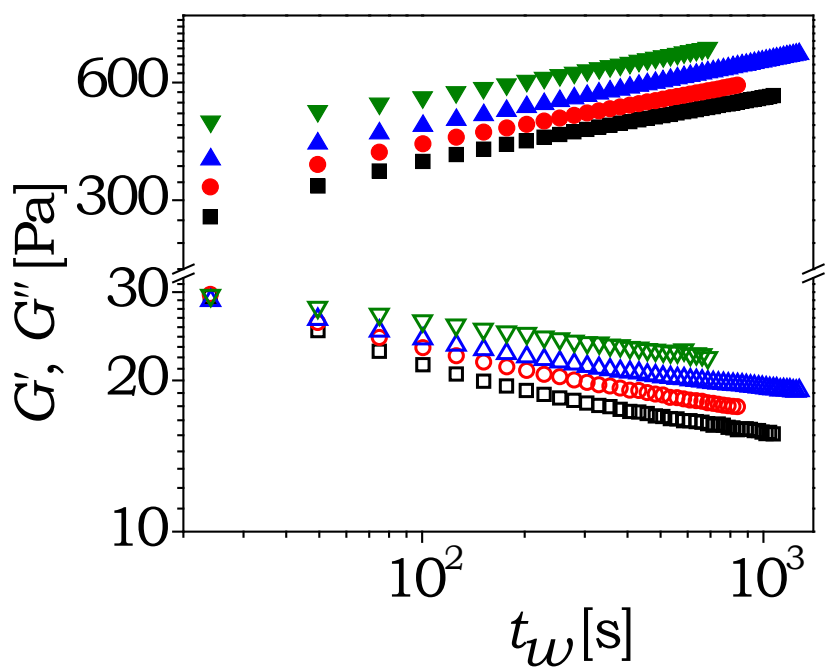

(f)

**Figure S1**. Evolution of storage (closed symbols) and loss (open symbols) modulus for 2.8 weight % no salt system having idle time $t_i$ = (a) 6, (b) 9, (c) 12, (d) 15, (e) 18 and (f) 60 days. (Black squares 10°C, Red circles 20°C, Blue up triangles: 30°C and Green down triangles 40°C).